\documentclass[reprint, nofootinbib, amsmath, amssymb, aps, pra]{revtex4-1}

\usepackage{graphicx}
\usepackage{dcolumn}
\usepackage{bm}
\usepackage[urlcolor=blue, hyperindex, colorlinks, bookmarks=true]{hyperref}
\begin{document}

\title{Assessing the Influence of Broadband Instrumentation Noise on Parametrically Modulated Superconducting Qubits}

\author{E. Schuyler Fried}
\thanks{These authors contributed equally to this work. }
\author{Prasahnt Sivarajah}%
\thanks{These authors contributed equally to this work. }
 \author{Nicolas Didier}
 \author{Eyob A. Sete}
 \author{Marcus P. da Silva}
 \author{Blake R. Johnson}
 \author{Colm A. Ryan}
\affiliation{%
 Rigetti Computing, 775 Heinz Avenue, Berkeley, 94710, CA
}%
\date{\today}
\begin{abstract}
With superconducting transmon qubits --- a promising platform for quantum information processing ---  two-qubit gates can be performed using AC signals to modulate a tunable transmon's frequency via magnetic flux through its SQUID loop. However, frequency tunablity introduces an additional dephasing mechanism from magnetic fluctuations. In this work, we experimentally study the contribution of instrumentation noise to flux instability and the resulting error rate of parametrically activated two-qubit gates. Specifically, we measure the qubit coherence time under flux modulation while injecting broadband noise through the flux control channel. We model the noise's effect using a dephasing rate model that matches well to the measured rates, and use it to prescribe a noise floor required to achieve a desired two-qubit gate infidelity. Finally, we demonstrate that low-pass filtering the AC signal used to drive two-qubit gates between the first and second harmonic frequencies can reduce qubit sensitivity to flux noise at the AC sweet spot (ACSS)~\cite{Didier2018}, confirming an earlier theoretical prediction. The framework we present to determine instrumentation noise floors required for high entangling two-qubit gate fidelity should be extensible to other quantum information processing systems.
\end{abstract}

\maketitle

\section{\label{sec:intro}Introduction}
In recent years, superconducting transmon-qubit based quantum computers have decreased their error rates to levels reaching $\le 1\%$~\cite{Barends2014, Rol2019, Sheldon2016}, close to the thresholds required for fault tolerance~\cite{Aharonov1997, Kitaev1997, Knill1998, Raussendorf2007, Knill2005}. Two-qubit error rates, in particular, have made notable strides in the last decade. One way to implement two-qubit gates on the transmon platform is to utilize coupling between a fixed and tunable transmon, separated in frequency space, and modulate the tunable transmon's frequency with an AC signal to satisfy a resonance condition between the qubits~\cite{Caldwell2018}. Separating the coupled qubits' frequencies effectively reduces the always-on coupling between them to a negligible level. However, the use of a frequency-tunable transmon presents an additional dephasing mechanism that arises from fluctuations of the transmon frequency via coupling to the magnetic environment~\cite{Luthi2018}. This can be mitigated by biasing the transmon at a DC magnetic flux offset where the qubit frequency is first-order insensitive to flux noise, herein referred to as the ``DC sweet spot" (DCSS)~\cite{Vion2002}. Yet, several two-qubit gate schemes require modulating the transmon away from this offset during gate operation, thereby reintroducing its sensitivity to flux noise~\cite{Reagor2018, Barends2014, Rol2019, Chu2019}. Recently, Didier et al.~\cite{Didier2018}~proposed a way to overcome this using a parametric modulation scheme wherein an AC signal is used to drive the tunable transmon to an ``AC sweet spot" (ACSS), at which its average frequency is insensitive to $1/f$-type flux noise. By using the ACSS, Hong et al.~\cite{Hong2019}~were later able to demonstrate a recovery of the coherence time to the level observed without modulation, and used this operating point to implement a two-qubit CZ gate with error rate $\approx 1\%$. This work also briefly touched on the influence of improved instrumentation. In particular, having eliminated qubit susceptibility to $1/f$-type flux noise, broadband noise becomes the dominant contributor to dephasing at the ACSS. This noise stems from either background or control electromagnetic fields coupling to the qubit. While the former is determined by defects on the device and other environmental coupling, the latter is determined by the instrument noise floor and control line attenuation. As the parametric two-qubit gate scheme continues to drive towards lower error rates, it becomes imperative to (1) understand the relative contribution of the background and control noise sources and (2) provide concrete instrument noise floor requirements to achieve a desired gate infidelity. 

Here, we experimentally address these goals by utilizing an artificial broadband noise source to study dephasing at the ACSS. We begin in Section~\ref{section2} by first assessing the noise source using a Carr-Purcell-Meiboom-Gill (CPMG) sequence~\cite{Meiboom1958}. More specifically, the tunable transmon, driven with the CPMG pulse sequence, is used as a spectrometer to validate the broadband nature of our noise source. In Section~\ref{section3}, we then use this noise source to study the coherence time at the ACSS as a function of the noise power spectral density (PSD).  We model the effect using an additive rate-equation model for dephasing that agrees well with our data. In Section~\ref{section4}, we then extend this model to formulate a noise floor requirement to achieve a desired two-qubit gate infidelity at the ACSS. Finally, in Section~\ref{section5} we experimentally demonstrate a way to mitigate the effect of the broadband noise by filtering the AC signal above its fundamental frequency.

\section{Broadband noise source}\label{section2}

\begin{figure}
    \includegraphics[width=\columnwidth]{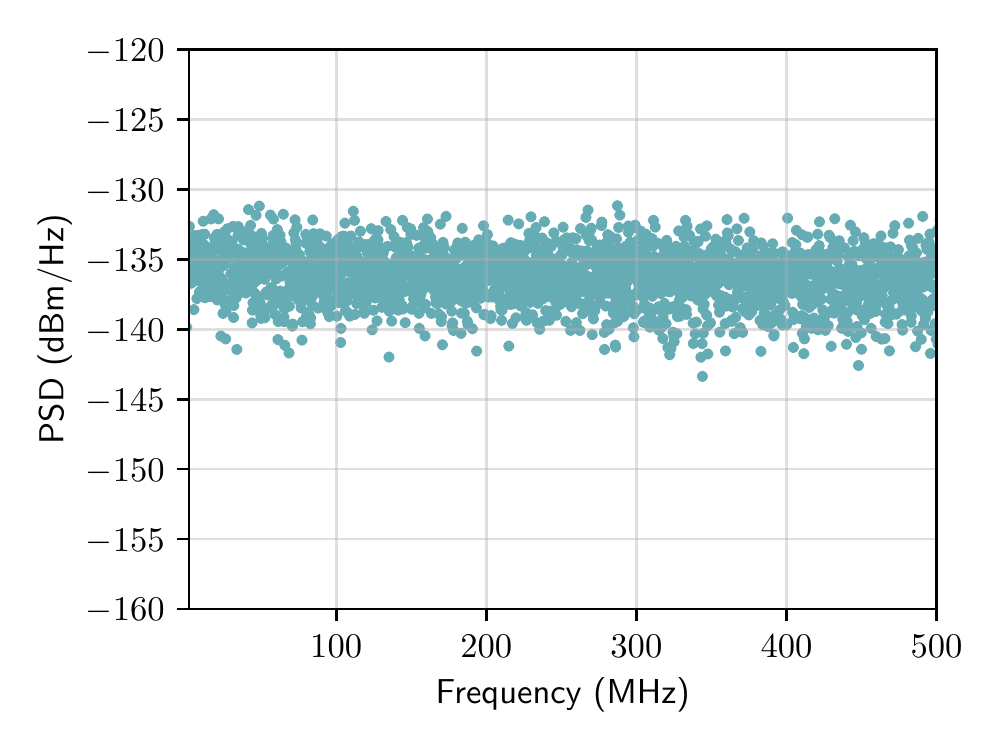}
    \caption{The noise PSD of the TEK as measured on a Agilent  N9020A  MXA  Signal Analyzer. The spectrum has a near-negligible slope of $-8 \times 10^{-9}\,$dBm/Hz$^2$, confirming that the noise generated is broadband at room temperature up to $500\,$MHz. Note that the plotted values do not include corrections for the Agilent  N9020A  MXA  Signal Analyzer noise floor.}
    \label{fig:tek_spec}
\end{figure}

To measure a qubit's broadband magnetic flux noise sensitivity, we require an instrument that can generate sufficiently broadband noise of varying noise power. We use the qualifier ``sufficiently" since no noise source is infinitely broadband. Sufficiency means possessing a flat noise spectrum in the frequency regime $\omega/2\pi<1/t \sim 100\,$MHz, where $t > 10\,$ns is the typical time-resolution dictated by the single-qubit gate durations. To this end, we chose the Tektronix Arbitary Waveform Generator 5208 (TEK). The TEK offers a noise function generator that possesses a user configurable noise PSD and a flat noise spectral density out to $\omega  /2 \pi < 500$\,MHz (see Fig.~\ref{fig:tek_spec}). Herein, the reported noise PSD $S^{\text{inj}}_{\text{noise}}$ is the total noise injected through the flux signal chain, as measured at room temperature by directly connecting the voltage or current source used to control magentic flux to an Agilent N9020A MXA Signal Analyzer and correcting for the N9020A's internal noise floor. Note that when only utilizing the TEK, the minimum value of  $S^{\text{inj}}_{\text{noise}}$ will equal $-142\,$dBm/Hz, i.e.~the native noise floor of the TEK. Later on, we will also combine a custom AWG signal as a part of the flux signal chain. The resulting minimum value of $S^{\text{inj}}_{\text{noise}}$ will be higher due to the additional broadband noise from the custom AWG \footnote{Note that the Rigetti custom AWG has a native noise floor of $-143\,~$dBm/Hz, but it rises to $-135\,~$dBm/Hz with its DAC at full scale.}. This will ensure that we can prescribe a practically measurable noise requirement. 

\par\bigskip
\begin{figure}
    \includegraphics[width=\columnwidth]{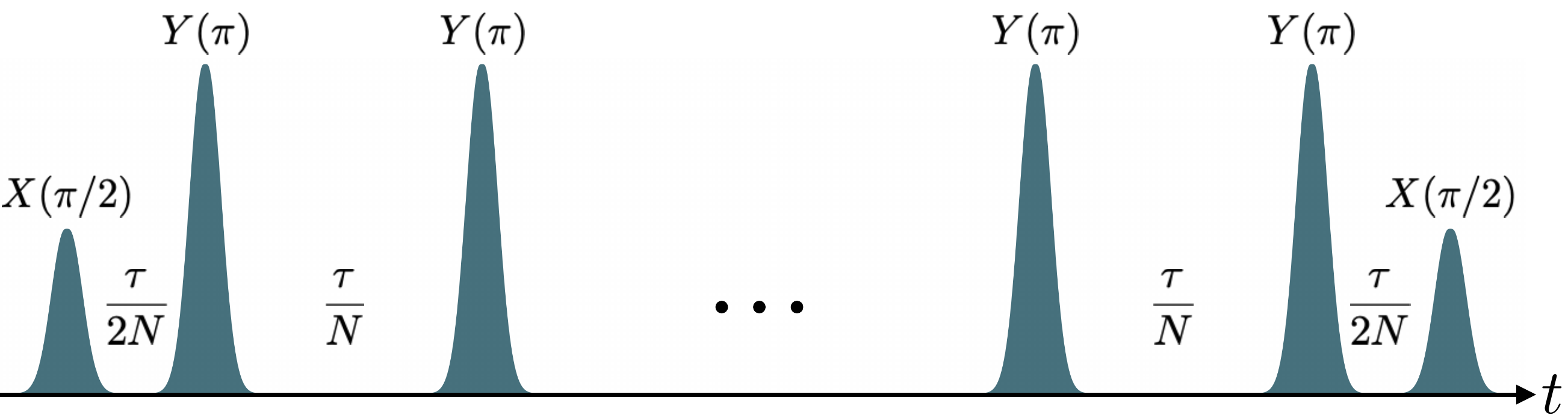}
    \caption{We use a transmon and a CPMG pulse sequence of length $\tau$ and $N$ pulses as a spectrometer to confirm the flux signal chain provides a broadband noise spectrum as sampled by the qubit.}
    \label{fig:cpmg_seq}
\end{figure}

\begin{table}
\centering
\begin{ruledtabular}
\begin{tabular}{cccc}
\hline
\textbf{Qubit} & \boldsymbol{$\mathrm{f}_{01}^{\mathrm{max}}\,$}(MHz) & \boldsymbol{$\mathrm{f}_{01}^{\mathrm{min}}\,$}(MHz) & \boldsymbol{$\mathrm{Anh.}$} (MHz) \\
\hline
Transmon $1$ & $4953$ & $4303$ & $-195$  \\
\hline
Transmon $2$ & $4649$ & $3878$ & $-188$  \\
\hline
Transmon $3$ & $4764$ & $4115$ & $-188$  \\
\hline
\end{tabular}
\end{ruledtabular}
\caption{\label{tab:1} Measured parameters for the tunable transmon qubits used in the experiment.}
\end{table}
   
To verify the instrument possesses a flat noise spectrum, we utilize a CPMG measurement and the transmon together as a spectrometer~\cite{Bylander2011}. The CPMG measurement consists of a Ramsey  sequence  with $N$ $\pi$ echo pulses inserted during the free evolution time $\tau$ (see Fig.~\ref{fig:cpmg_seq}). The measurement can be thought of as a narrow-band filter for longitudinal ($\sigma_{z}$) noise~\cite{Bylander2011, Biercuk2011, Green2012, OMalley2016}. Since magnetic flux coupling is longitudinal, CPMG serves as an excellent measure of its noise spectrum. Additionally, by measuring the transmon with a CPMG sequence, we confirm that the signal chain does not reshape the noise spectrum observed at room temperature. The dependence of the pure dephasing time $T_\phi$ on the filter function can be described by the formula
\begin{align}\label{eqn1}
    1/T_\phi^N = \left(\frac{d\omega}{d\Phi}\right)^2\int d\omega~S(\omega)f(\omega, N),
\end{align}
where $d\omega / d\Phi$ is the sensitivity of the tunable qubit frequency to flux, $S(\omega)$ is the flux noise PSD, and $f(\omega, N)$ is the CPMG filter function for $N$ echo pulses. The frequency dependence of the filter is given by~\cite{Bylander2011, OMalley2016}~
\begin{align}\label{eqn2}
    f(\omega, N) = \tan^2\left(\frac{\omega t}{4N}\right)\frac{\sin^2(\omega t / 2)}{(\omega / 2)^2}.
\end{align}
As $N$ increases for a fixed $\tau$, the filter function peaks at successively higher frequencies. For purely broadband noise, $N$ does not influence $T_{\phi}^N$ since the net integrated noise is the same regardless of the center frequency of the filter. Conversely, for noise with a frequency roll-off (e.g. $1/f$), increasing $N$ leads to a reduction in the net integrated noise (i.e.~$T_\phi^0 < T_{\phi}^1 < ...)$. This holds true until $f(\omega, N)$ is peaked at a frequency where broadband noise dominates (e.g. $1/f$ noise-corner), making the generic assumption that our noise is a sum of broadband and $1/f$ type noise. Thus, our aim is to successively increase the broadband noise from our instrument until it becomes the dominant contributor. If the instrument noise is sufficiently broadband, we should expect to see $T_\phi^N$ converge for all values of $N$.

\begin{figure}
    \includegraphics[width=\columnwidth]{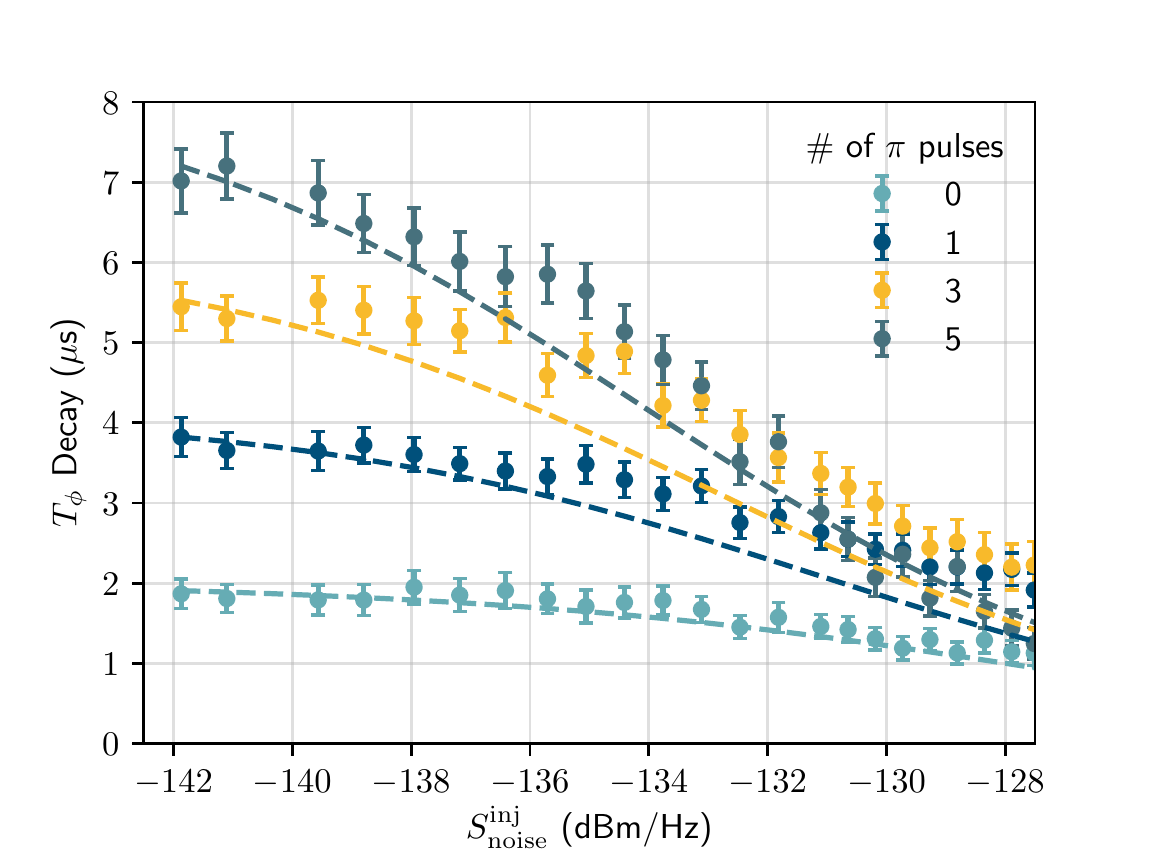}
    \caption{Experimentally measured (points) and numerically calculated (dashed lines) dephasing times for tunable transmon $1$ (ref. Table \ref{tab:1}) parked at $0.282\Phi_0$ as a function of varying number of CPMG echo pulses $N$ and flux noise PSD $S^{\text{inj}}_{\text{noise}}$ injected from the TEK. Numerical calculations were done using Eqs.~\eqref{eqn1} and~\eqref{eqn2}.  The convergence of the dephasing time with increasing $S^{\text{inj}}_{\text{noise}}$ for different $N$ demonstrates the broadband spectrum of the injecting noise source. The numerical calculation assumes a static background $1/f$ noise spectrum and a varying broadband noise power.}
    \label{fig:tphi_convergence}
\end{figure}

To setup our experiments, we first bias a tunable transmon away from the DCSS such that it has a first order non-zero dependence on flux noise (i.e.~$d\omega /d\Phi\neq 0$). This ensures sufficient sensitivity of $T_\phi$ to our injected noise. We then measure $T_{\phi}^{N}$ for a range of $N$ and range of  injected broadband noise $S^{\text{inj}}_{\text{noise}}$ from the TEK \footnote{Note that functionally, $T_{2}^{N}$ and $T_1$ are the relevant quantities measured during a CPMG experiment, which are then used to calculate $T_{\phi}^N = {1}/({1}/{T_2^N}-{1}/{(2T_1)})$.}. As shown in Fig.~\ref{fig:tphi_convergence}, we find that increasing $N$ also increases $T_{\phi}^{N}$ when the colored background noise dominates. Conversely, $T_{\phi}^{N}$ converge for all $N$ when the injected broadband noise begins to dominate. This convergence is also confirmed by numerical calculations when using a combination of static $1/f$ noise and varying broadband noise, confirming the expected dependence on broadband noise.

\section{Broadband Noise at The AC Sweet Spot}\label{section3}

\begin{figure}
    \includegraphics[width=\columnwidth]{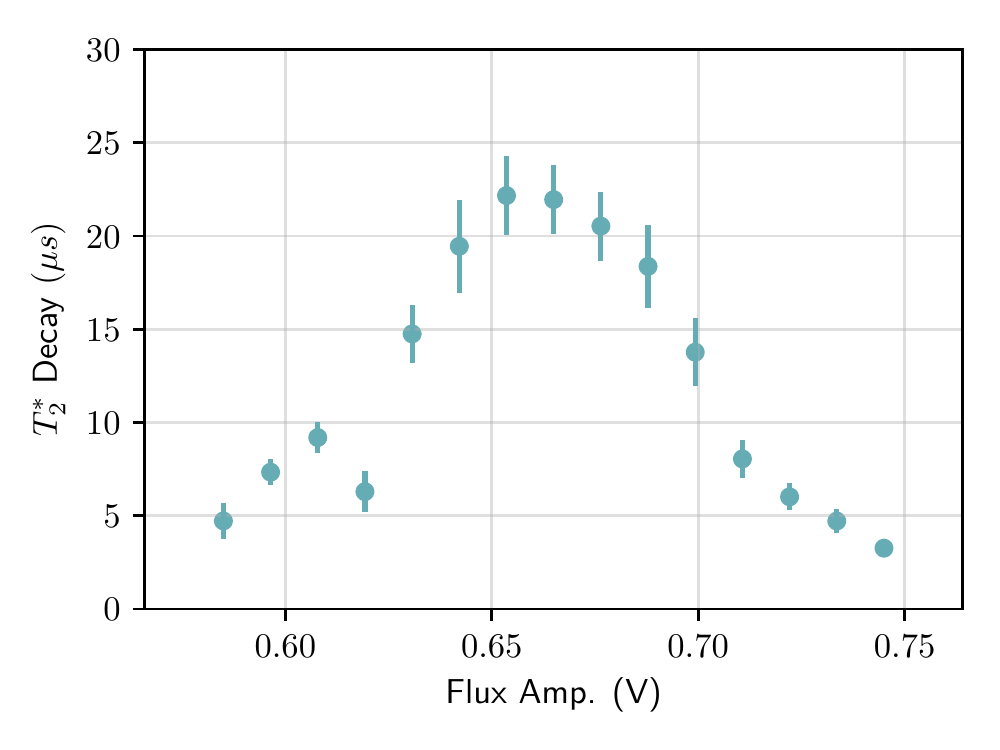}
        \caption{Tunable transmon $2$'s (ref. Table \ref{tab:1}) coherence times as a function of the AC flux pulse amplitude, measured by applying a $200\,~\mathrm{MHz}$ flux pulse during the Ramsey free evolution time. With no additional injected flux noise, the coherence time shows a recovery to values without AC modulation ($\sim26\,~\mu\mathrm{s}$) at $\sim0.67\,~\mathrm{V}$, demonstrating the existence of the ACSS.}
        \label{fig:acss_emerges}
\end{figure}

\begin{figure}
    \includegraphics[width=\columnwidth]{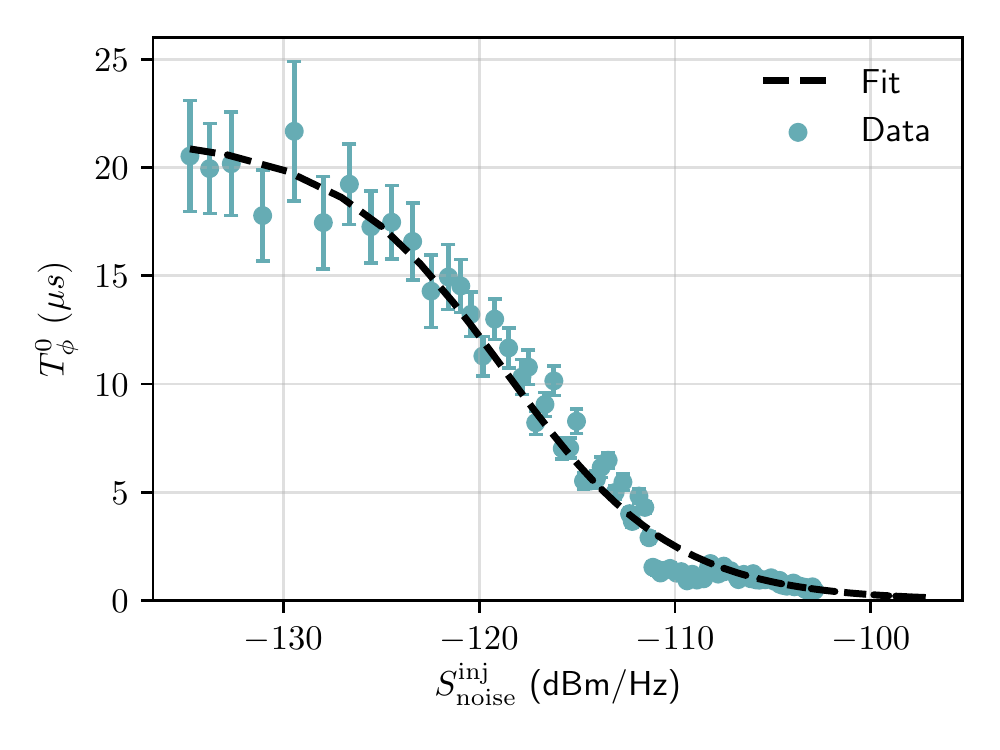}
    \caption{Tunable transmon $3$'s (ref. Table \ref{tab:1}) dephasing times under modulation as a function of the injected broadband noise PSD $S^{\text{inj}}_{\text{noise}}$ at a fixed AC flux pulse amplitude corresponding to the ACSS.}
    \label{fig:fit_at_acss}
\end{figure}
Having demonstrated that our instrument's noise spectrum is sufficiently broadband, our goal is to derive a relationship between the instrument broadband noise and the qubit coherence time at the ACSS. To this end, we combine the signals from TEK and a custom AWG as the qubit's flux source. The TEK handles DC flux and broadband noise generation, while the AWG handles AC flux signals. As such, the baseline value for $S^{\text{inj}}_{\text{noise}} = -135\,$dBm/Hz, set by the combined native noise floor of the TEK and custom AWG. 

To setup the experiment, we first ensure the qubit is biased to the DCSS. We then modulate to the ACSS using an AC signal from the AWG. In Fig.~\ref{fig:acss_emerges}, we show measurements of the Ramsey fringe decay time as a function of the AC signal amplitude under baseline injected noise. The measurements are taken by applying the flux pulse during the free evolution time of the $T_2^*$ measurement. The emergence of the ACSS is clearly observable at $\sim$0.67 V, where the gradient of the average frequency vanishes~\cite{Didier2018}. Operating at this amplitude, we then inject increasing levels of broadband noise $S^{\text{inj}}_\text{noise}$. For each $S^{\text{inj}}_\text{noise}$, we perform $T_1$ and $T_2^*$ Ramsey experiments and calculate $T_\phi(S^{\text{inj}}_\text{noise})$ \footnote{Note that we perform Ramsey instead of CPMG sequences because we are protected from frequency dependent flux noise at the ACSS.}. Our experimental results, shown in Fig.~\ref{fig:fit_at_acss}, display a clear dependence of qubit dephasing on the broadband noise PSD at the AC sweet spot. 

We model the transmon coherence time under modulation by assuming that there are two sources of noise -- background and control. The former arises from device defects and environmental couplings while the latter is determined by the injected instrument noise. The rate equation for $T_\phi(S^{\text{inj}}_\text{noise})$ can be written as
\begin{align}\label{eqn3}
    \frac{1}{T_\phi(S^{\text{inj}}_\text{noise})} = \frac{1}{T_\phi^{\text{bg}}} + \frac{1}{T_\phi^{\text{ctrl}}(S^{\text{inj}}_\text{noise})}
\end{align}
where $T_\phi^{\text{bg}}$ and $T_\phi^\text{ctrl}(S^{\text{inj}}_\text{noise})$ refer to the coherence time due to static background and injected instrument noise, respectively. Having shown in Section~\ref{section2} that our noise source generates broadband noise, the following relation therefore holds~\cite{OMalley2016, Didier2018}~
\begin{align}\label{eqn4}
    \frac{1}{T_\phi^{\text{ctrl}}(S^{\text{inj}}_\text{noise})}= \alpha S^{\text{inj}}_\text{noise},
\end{align}
where $\alpha$ accounts for both the sensitivity of the qubit frequency to drive amplitude and the attenuation from the instrument output (where $S^{\text{inj}}_\text{noise}$ is measured) to the SQUID loop. In Fig.~\ref{fig:fit_at_acss}, we fit the data to Eq.~\eqref{eqn3} with $\alpha$ as a free parameter and find excellent agreement to the data. 

With $\alpha$ and Eqs.~\eqref{eqn3} and~\eqref{eqn4} in hand, we have the necessary tools for relating the PSD measured on a spectrum analyzer $S^{\text{inj}}_\text{noise}$ to the resulting coherence time from this noise channel. In the next section, we utilize these tools to predict entangling gate infidelity due to a given instrument noise PSD.

\section{Converting AC Flux Noise to Entangling Gate Fidelity}\label{section4}

\begin{figure}
    \includegraphics[width=\columnwidth]{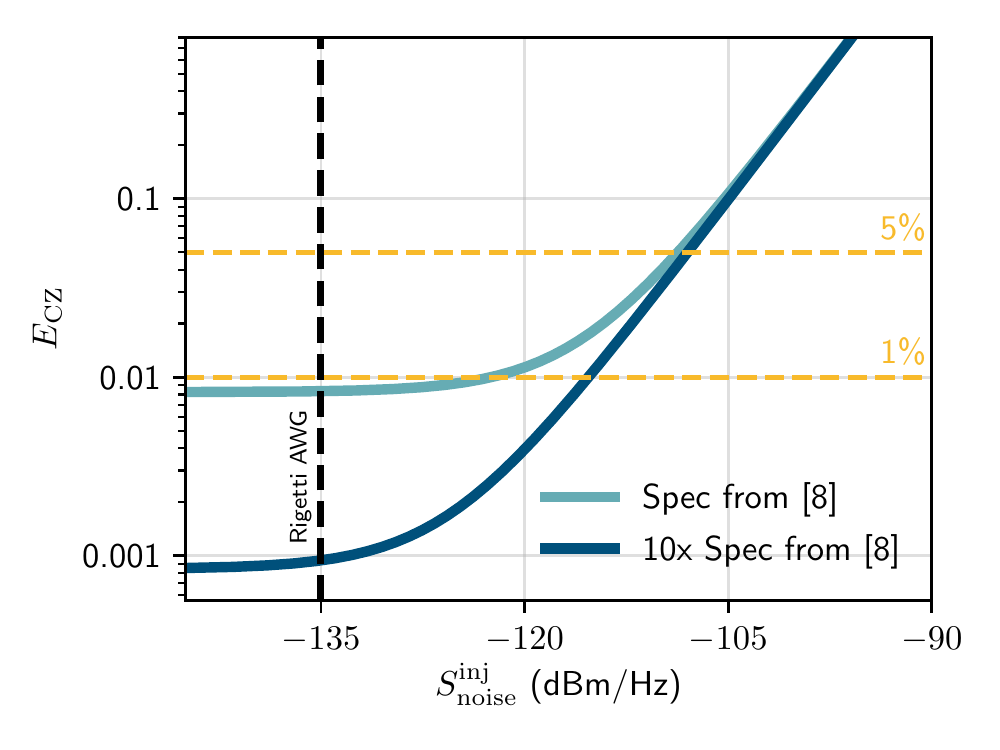}
    \caption{We can extrapolate the CZ infidelity as a function of the broadband noise PSD, given gate and coherence times from Table \ref{tab:5/tc} and the relation in Eq.~\eqref{eqn6}. We see the infidelity reach $1\%$ at $-123\,~$dBm/Hz noise PSD and saturate to $0.83\%$ at $-138\,~$dBm/Hz due to $T_1$ and $T_2$ coherence limits. By increasing the $T_1$ and $T_2$ tenfold, we can achieve $1\%$ infidelity at $-115\,~$dBm/Hz and saturate to $0.085\%$ infidelity at $-147\,~$dBm/Hz.}
    \label{fig:cz_infid}
\end{figure}

To determine the instrument noise floor needed to achieve a desired two-qubit gate infidelity, we use the model for relating the coherence time to parametric entangling gate fidelity established in Ref.~\cite{Didier2018}. In the limit where the entangling gate time $t_g \ll T_1, T_2$, we can estimate the coherence limited gate infidelity $E_{\mathrm{CZ}}$ using a simple analytical formula that includes the gate time and decay rates of the two qubits via the relation
\begin{align}\label{eqn5}
    E_{\mathrm{CZ}} \simeq 0.3\frac{t_g}{T_1^F} + 0.5
    \frac{t_g}{T_1^T} + 0.3625\frac{t_g}{T_\phi^F} + 0.7625\frac{t_g}{T_\phi^T}.
\end{align}
where $t_g$ refers to the gate time, and the superscript $F$ and $T$ refer to the fixed and tunable qubit. The relation holds for the CZ gate where nutation occurs between states $|11\rangle_{FT}$ and $|02\rangle_{FT}$. As such, we first test its validity by calculating the CZ infidelity for the gate parameters in the work of Hong et al.~\cite{Hong2019}~(see Table \ref{tab:5/tc}). We find agreement to within $1\%$ of the observed experimental infidelity, lending credence to the use of Eq.~\eqref{eqn5}.
\begin{table}
\centering
\begin{tabular}{lccc}
\toprule
\hline
\textbf{Qubit} & $T_1$ & $T_\phi$\\
\hline
\colrule
Fixed ($\mu$s)& 17.9 & 28.3  \\
\hline
Tunable ($\mu$s) & 25.3 & 35.0  \\
\hline
$t_g$ (ns) & \multicolumn{2}{c}{176 }\\
\hline
$F_{\mathrm{CZ}}$ (Meas.) & \multicolumn{2}{c}{98.8\% }\\
\hline
$F_{\mathrm{CZ}}$ (Coh. Lim.) & \multicolumn{2}{c}{99.1\% }\\
\hline
\botrule
\end{tabular}
\caption{\label{tab:5/tc}Gate and coherence times under AC modulation for the tunable and fixed qubit pair in~\cite{Hong2019}~used to demonstrate $\approx $1\% CZ gate infidelity.}
\end{table}
We then make use of Eqs.~\eqref{eqn3} and~\eqref{eqn4} to re-write Eq.~\eqref{eqn5} as
\begin{align}\label{eqn6}
    E_{\mathrm{CZ}}(S^{\text{inj}}_\text{noise}) &\simeq 0.3\frac{t_g}{T_1^F} + 0.5\frac{t_g}{T_1^T} + 0.3625\frac{t_g}{T_\phi^F} \nonumber \\ &+ 0.7625t_g\left(\frac{1}{T_\phi^{T, \text{bg}}} + \alpha S^{\text{inj}}_\text{noise}\right),
\end{align}
Using the parameters from Table \ref{tab:5/tc} (assuming that $T_\phi^{T, \text{bg}}$ refers to the $T_\phi^{T}$ measured without injected noise) and $\alpha$ from the data fitted in Fig.~\ref{fig:fit_at_acss},  we plot the expected gate infidelity as a function of a given noise PSD in Fig.~\ref{fig:cz_infid}.

From this data, we conclude that a noise floor of $-123\,~$dBm/Hz is sufficient for $1\%$ CZ gate infidelity. Due to our attenuation scheme, this translates to roughly $-143\,~$dBm/Hz at the bottom of the fridge. Given that the custom AWG utilized in our work and~\cite{Hong2019}~has a native noise floor of less than $-140\,~$dBm/Hz and a noise floor of $-135\,~$dBm/Hz with the DAC at full scale, we conclude that further improvements to gate infidelity are not limited by the instrument noise floor but rather the background noise.

\section{Filtering AC Flux Noise}\label{section5}

In the presence of strong broadband flux noise, it is clear that the ACSS vanishes (see Fig.~\ref{fig:fit_at_acss}). This is in part because the response of the qubit to flux modulation is highly nonlinear and so the qubit frequency $\omega$ oscillates at many harmonics of the AC signal. The effect is modelled as a time-dependent qubit frequency $\omega(t)$ given as a Fourier series
\begin{align}\label{eqn7}
    \omega(t) = \sum_k\omega_k \cos[k(\omega_m t + \theta_m)]
\end{align}
where $\omega_k$ refers to Fourier coefficient, $\omega_m$ is the AC signal modulation frequency, and $\theta_m$ is the modulation phase. While the slope of the fundamental harmonic $d\omega_0/d\Phi$ with respect to magnetic flux vanishes at the ACSS, higher harmonics with $k\ge 2$ have a non-zero slope at this point. As such, Didier et al.~proposed eliminating the susceptibility of dephasing to these harmonics by placing a low-pass filter between $\omega_m$ and $2\omega_m$~\cite{Didier2018}. 

\begin{figure}
    \includegraphics[width=\columnwidth]{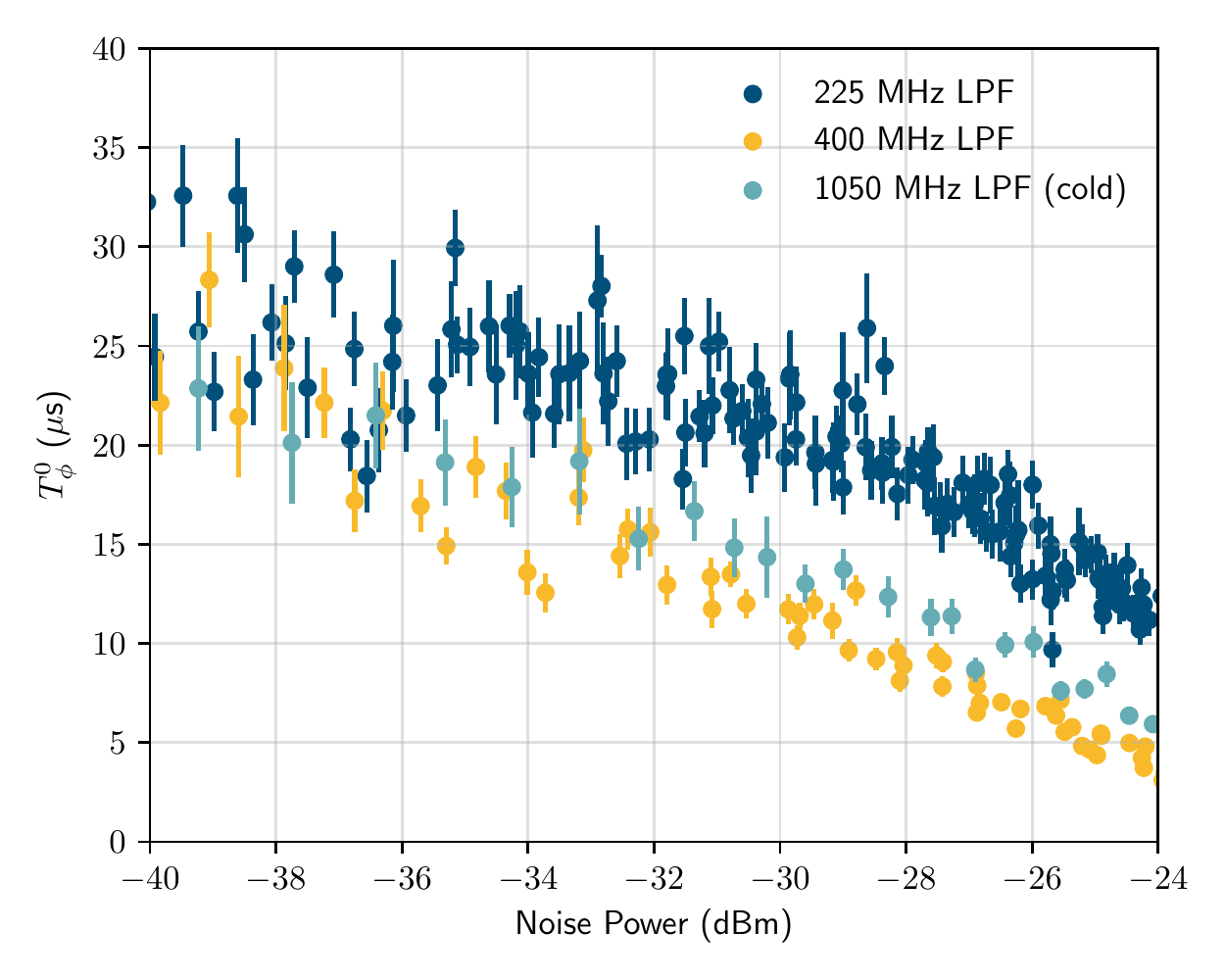}
    \caption{Measured coherence times of tunable transmon $3$ at the AC sweet spot as a function of broadband noise power with varying filter configurations and an AC flux pulse at 200 MHz. Placing a filter between 200 and 400 MHz shows a dramatric improvement in coherence times.}
    \label{fig:filter_fig}
\end{figure}

To experimentally verify the effect, we repeated the measurement of $T_{\phi}(S^{\text{inj}}_\text{noise})$ at the AC sweet spot as was done in Fig.~\ref{fig:fit_at_acss}, but now with a low pass filter on the flux line at room temperature. To also account for the differing noise powers for the filtration schemes, we show in Fig.~\ref{fig:filter_fig} the results as a function of total noise power instead of power spectral density. We find that the dephasing time $T_\phi$ is unchanged when the filter is placed above the first harmonic, but it is greatly increased if the filter roll off is placed between the fundamental and the first harmonic. This confirms that having eliminated the dephasing contribution from the fundamental AC pulse frequency at the ACSS, noise at the second harmonic is the next leading contributor to qubit dephasing under modulation. It also confirms that this problem has a tractable solution: it can be eliminated by the use of passive filtration, assuming the noise is coming down the signal line. 

\section{Conclusion}

We have validated a noise generator's broadband spectrum and calculated a qubit coherence time from noise PSD. Using our validated noise generator, we developed and presented a framework for estimating the parametric entangling gate fidelity from the broadband noise PSD on a tunable transmon qubit's flux control line. Finally, we employed low pass filters as a means to reducing a tunable qubit's sensitivity to AC flux noise. We hope this framework can be used to inform requirements on flux delivery instruments needed to achieve high entangling gate fidelity and facilitate future studies on noise level requirements for other control instruments. 

This work was funded by Rigetti \& Co Inc., dba Rigetti Computing. We thank Deanna Abrams, Sabrina Hong, and Alexander Papageorge for comments and useful discussions, the Rigetti quantum software team for providing tooling support, the Rigetti fabrication team for manufacturing the device, the Rigetti technical operations team for fridge maintenance, the Rigetti cryogenic hardware team for providing the chip packaging, and the Rigetti control systems and embedded software teams for creating the Rigetti custom AWGs.

\end{document}